\documentclass[prd,aps,showpacs,letterpaper]{revtex4}
\usepackage{amsmath}
\usepackage{graphicx}
\usepackage{amssymb}
\usepackage{epsfig}
\usepackage{hyperref}

\def\tal{{\widetilde \alpha}}
\def\tM{{\widetilde M}}

\long\def\symbolfootnote[#1]#2{\begingroup%
\def\thefootnote{\fnsymbol{footnote}}\footnote[#1]{#2}\endgroup}

\begin{document}
\title{Thermodynamics of static black objects in $D$ dimensional Einstein-Gauss-Bonnet gravity with $D-4$ compact
dimensions}
\author{C. Sahabandu\symbolfootnote[1]{chetiya@physics.uc.edu}}
\author{P. Suranyi\symbolfootnote[2]{suranyi@physics.uc.edu}}
\author{C. Vaz\symbolfootnote[3]{vaz@physics.uc.edu}}
\author{L. C. R. Wijewardhana\symbolfootnote[4]{rohana@physics.uc.edu}}
\affiliation{Department  of Physics,
University of Cincinnati, Cincinnati, Ohio 45221}
%\date{12.8.2003}

\begin{abstract}
We investigate the thermodynamics of static black objects such as black holes, black
strings and their generalizations to $D$ dimensions (``black branes") in a gravitational theory
containing the four dimensional Gauss-Bonnet term in the action, with $D-4$ dimensions compactified torus. The entropies of black holes and black branes are compared to obtain information
on the stability of these objects and to find their phase diagrams. We demonstrate the existence of a critical
mass, which depends on the scale of the compactified dimensions, below which the black hole entropy dominates
over the entropy of the black membrane.
\end{abstract}
\maketitle

\section{Introduction}
In dimension $D\geq 5$, the gravitational action may be modified to include higher order curvature terms
while keeping the equations of motion to second order, provided the higher order terms appear in specific
combinations corresponding to the Gauss-Bonnet invariants of even dimensions $d\leq D$ \cite{lovelock}.
String theory predicts that such Gauss-Bonnet terms arise as higher order corrections in the
heterotic string effective action. The simplest of these terms is the four dimensional Gauss-Bonnet
term. It arises as the next to leading order correction to the gravitational effective action in
string theory \cite{zwei,zumi} and has recently been receiving considerable attention as a gravitational
alternative to dark energy \cite{odintsov}.

In this paper we shall be concerned with compact black objects in such theories. Boulware and Deser (BD)
found exact black hole solutions~\cite{boulware} in a gravitational theory with a four dimensional
Gauss-Bonnet term modifying the usual Einstein-Hilbert action.  These solutions are generalizations of
the $D$-dimensional black hole solutions found a long time ago by Tangherlini~\cite{tangherlini}. Under
certain conditions, a single horizon enclosing a spacelike singularity exists and the global topology of
the manifold is identical to that of a Schwarzschild black hole. The thermodynamics of BD black holes was
subsequently discussed by Myers and Simon~\cite{myers} and by Wiltshire~ \cite{wiltshire}. They started from the observation of Bardeen,
Carter and Hawking~\cite{carter,bardeen} that the surface gravity is constant and inversely proportional
to the periodicity in imaginary time of the metric, thus relating the temperature of the black
hole to it. To construct a candidate entropy, they employed the same thermodynamic identities which
appear in Einstein Gravity \cite{carter2}, identifying the Euclidean action with the free energy. They
then derived an expression for the entropy of a black hole in Gauss-Bonnet gravity, concluding that
the entropy is not simply proportional to the area of the horizon of the black hole as it is in Einstein
gravity but rather has an extra term proportional to the Gauss-Bonnet coupling parameter.
Though it is a fairly natural identification to make, there is still no proof that the entropy so defined
obeys the second law ($dS \geq 0$) as this depends on the dynamics. Due to its interest in string theory,
there have subsequently been many works studying the entropy of black holes in spherically symmetric Gauss
Bonnet and Lovelock gravity with positive an negative cosmological constant~ \cite{bhent}, although a
convincing treatment of the second law remains elusive in all cases.

If some of the extra dimensions are compactified, then in addition to black hole solutions, black
string solutions (or ``black membrane" solutions) also exist.  While in Einstein gravity these are trivial
extensions of black hole solutions in extra dimensions, it has been pointed out by Kobayashi and
Tanaka~\cite{kobayashi} that the solution has to be modified and an extra asymptotic charge, in addition
to the Arnowitt-Deser-Misner (ADM) mass~ \cite{adm}, appears in the presence of a Gauss-Bonnet term. In the current paper we set out to
discuss the thermodynamic properties of black holes and the generalization of
black strings to $D$ dimensions in Einstein-Gauss-Bonnet gravity, with $D-4$ compact dimensions. We compare
the entropies of various black objects, which allows us, assuming that the second law of thermodynamics
holds, to investigate their stability.

In Sec. 2 we summarize our knowledge of the thermodynamics of black holes in Gauss-Bonnet gravity for
$D\ge5$. In Sec. 3. we discuss black strings and their
generalization to $D\ge6$ ``black branes." In Sec. 4 we will compare the entropies of black holes and
black strings and construct a rough phase diagram for stable black objects.  We conclude the paper and
summarize our results in Sec. 5. In the Appendices we list the equations of motion and the coefficients of
the $\alpha-$expansion of the metric tensor up to third order.

\section{Black holes in Einstein-Gauss-Bonnet gravity and compactification}

The action of Einstein-Gauss-Bonnet gravity is\cite{Weinberg}
\begin{equation}
S=-\frac{1}{16\pi G_D}\int d^Dx\sqrt{-g}\left(R+\alpha L_{\rm GB}\right),
\label{action}
\end{equation}
where $R$ is the Ricci scalar, and
\begin{equation}
L_{\rm GB}=R^2-4R_{ab}R^{ab}+R_{abcd}R^{abcd}
\label{gauss}
\end{equation}
is the Gauss-Bonnet term. While the Gauss-Bonnet term is topological in $D=4$, in $D>4$ $L_{\rm GB}$ is
a scalar, modifying the action in a non-trivial manner.

The variation of the action in (\ref{action}) results in a modified Einstein equation of the form
\begin{equation}
G_{ab}-\frac{\alpha}{2}H_{ab}=0,
\label{einstein}
\end{equation}
where $G_{ab}=R_{ab}-g_{ab}R/2$ is the Einstein tensor and
\begin{equation}
H_{ab}=L_{\rm GB}g_{ab}-4RR_{ab}+8R_{ac}{R^c}_b+8R_{acbd}R^{cd}-4R_{acde}{R_b}^{cde}.
\label{lanczos}
\end{equation}
is the Lanczos tensor. Boulware and Deser have found exact, spherically symmetric, static solutions
of (\ref{einstein}) at every $D\ge5$. These have the form
\begin{equation}
ds^2=-f(r) dt^2+\frac{1}{f(r)}dr^2+r^2d\Omega_{D-2},
\label{bw-metric}
\end{equation}
where
\begin{equation}
f(r)=1+\frac{r^2}{2\tal}\left(1-\sqrt{1+\frac{4\tal\tM}{r^{D-1}}}\right).
\end{equation}
The parameters ${\tal}$ amd $\tM$ are defined as follows:
\begin{equation}
\tal=(D-4)(D-3)\alpha
\label{talpha}
\end{equation}
and
\begin{equation}
\tM=\frac{16 \pi G_DM}{(D-2)\Omega_{D-2}},
\label{tM}
\end{equation}
where
\begin{equation}
\Omega_{D-2}=\frac{2 \pi^{(D-1)/2}}{\Gamma((D-1)/2)}.
\end{equation}
is the volume of the $D-2$ dimensional unit sphere and $M$ is the ADM mass.

The dimensions $D=5$ and $D=6$ are somewhat special.  Solving the equation for the horizon $f(r)=0$
one finds a mass gap for black holes at $D=5$. No black holes exist unless $M>3\alpha/(4\pi G_5)$.
$D=6$ black holes also have a mass gap, but for an entirely different reason. In $D=6$, small
black holes are unstable due to a runaway mode~\cite{stability}.

The temperature of the black hole is obtained directly from the periodicity in imaginary time
of the lapse function. One rotates to Euclidean time ($t\to i\tau$) and produces a smooth Euclidean
manifold by identifying $\tau$ with perdiodicity $\beta$. This periodicity is reflected in the
Euclidean propagator of any quantum field propagating in the background spacetime and may be interpreted
as indicating that the fields are in thermal equilibrium with a reservoir at temperature $T=1/\beta$.
One finds that the periodicity is $\beta=2\pi/\kappa$, where $\kappa$ is the surface gravity of the event
horizon. For the black hole under consideration, the temperature as calculated by Myers and Simon
\cite{myers} and by Wiltshire~\cite{wiltshire} is
\begin{equation}
T=\frac{D-3}{4\pi r_h}\left[\frac{r_h^2 +\tal \frac{D-5}{D-3}}{r_h^2 + 2\tal}\right]
\end{equation}
where $r_h$ is the horizon radius, satisfying $f(r_h)=0$. Using the standard relationship
between the temperature and the entropy
\begin{equation}
S = \int \frac{d M}{T} + S_0
\label{entT}
\end{equation}
where $S_0$ is mass independent, the expression for the entropy of the spherical black
hole in Einstein-Gauss-Bonnet gravity becomes
\begin{equation}
S=\frac{\Omega_{D-2}}{4G_D}r_h^{D-2}\left(1+\frac{(D-2)}{D-4}\frac{2\tal}{r_h^2}\right),
\label{entsph}
\end{equation}
where following Myers and Simon~\cite{myers} $S_0$ is set to zero. This differs from the
expression of entropy in Einstein gravity, in which the entropy is proportional to the area of
the event horizon.

Below we consider black membranes. In what follows, we will assume that the temperature is also given by
\begin{equation}
T=\frac{\kappa}{2\pi},
\label{temperature}
\end{equation}
where $\kappa$ is the surface gravity of the membrane.  Further, we determine the entropy from
(\ref{temperature}) using (\ref{entT}) and leave the second law of black hole thermodynamics for
future consideration.

\section{Black branes}

Recently, Kobayashi and Tanaka~\cite{kobayashi} (KT) have investigated black string solutions for
$D=5$ of Einstein-Gauss-Bonnet gravity with one spatial dimension compactified on a circle. However, if one keeps
future phenomenological applications in  mind the case $D\geq 6$ is more interesting as the $D=5$
Arkani-Hamed - Dimopoulos - Dvali (ADD) model~\cite{add} of large extra dimensions modifies Newtonian
gravity at unacceptably large distances.  Therefore, one of our aims has been to extend KT to $D\ge6$.
For simplicity we will choose a compactification on a symmetric torus $(S^1)^{D-4}$. An unfortunate
consequence of our more general approach is our inability to employ numerical methods for the
investigation of the phase structure in the space of solutions.  We will see, however, that an expansion
in terms of the ratio $\sqrt{\alpha}L^{D-4}/G_{D}M$, also proposed in KT, provides respectable results
at all $D$, where $L$ is the length associated with each toroidal dimension.
KT have shown that, unlike in Einstein gravity, black string solutions acquire a non-trivial second
charge, $Q$,  in addition to a mass parameter, $M$. The asymptotic behavior of the
metric components is
\begin{eqnarray}
g_{tt}&=&1-\frac{2G_D M}{L^{D-4}r}+O(r^{-2})~,\label{metric2}\\
g_{ww}&=&1+\frac{G_D Q}{L^{D-4}r}+O(r^{-2})~.
\label{metric3}
\end{eqnarray}
where $G_D$ is Newton's constant in $D$ dimensions. For black string solutions (solutions with an event horizon)
the two charges have a unique relationship that can be studied either numerically, or, by employing analytic
expansion methods. We consider the explicit forms of these quantities in (\ref{ctcw}). For the metric describing a
black membrane in $D$ dimensions, $D-4$ of which are compactified on a torus, we use the ansatz
\begin{equation}
ds^2=-e^{2A(r)}dt^2+e^{2B(r)}dr^2+r^2d\Omega^2+e^{2\phi(r)}\sum_{i=1}^{D-4} dw_i^2.
\label{metric1}
\end{equation}
To derive the equations for the components of the metric for general $D$ we utilize the fact that the
equations are polynomial in $D$. These polynomials appear from summations over $D-k$ terms, with $k$
independent of $D$. As the calculation of the Einstein tensor requires at most two summations, the
resulting expression is at most quadratic in $D$.  Following similar considerations, we can conclude that
the Lanczos tensor is at most a quartic function of $D$. The coefficients of various terms can than be
identified by deriving these tensors at a few fixed values of $D$. We list the nonzero components of the
Einstein and Lanczos tensors for arbitrary $D$ in Appendix A.

Equation (\ref{einstein}) has no known exact black brane solutions. However, one can study the black
brane solutions in several different ways. Following KT we employ expansions around the horizon and in
the dimensionless variable $\sqrt{\alpha}/G_{4}M$.

\subsection{Horizon expansion}

One can investigate black string solutions further by expanding the metric in a variable $x=r-r_h$,
where $r_h$ is the radius of the horizon~\cite{kobayashi}.  This investigation yields the result that in $D=5$
the radius of the black string horizon has a lower bound $R=\sqrt{8\alpha}$. The generalization of this result to $D$ dimensions using the form of the Einstein and Lanczos tensors of Appendix A is straightforward.
We write the components of the metric~\cite{kobayashi} as
\begin{eqnarray}
e^{2A(r)}&=&a_1x+\frac{1}{2}a_2x^2+...\label{a-eq}\\
e^{-2B(r)}&=&b_1x+\frac{1}{2}b_2x^2+...\label{b-eq}\\
e^{2\phi(r)}&=&\phi_0+\phi_1x+\frac{1}{2}\phi_2x^2+...\label{phi-eq}
\end{eqnarray}
Substituting these expressions into (\ref{einstein}) and solving for the coefficients provides the following
expression for the leading coefficients (the coefficient $a_1$ and $\phi_1$ are undetermined)
\begin{eqnarray*}
b_1&=&\frac{(D-2)+\frac{8(D-2)}{D-4}\xi-48(D-6)\xi^2\pm(1+4\xi) \sqrt{(D-2)^2-48(D-2)^2\xi^2-1152(D-4)\xi^3}}
{4r_h \xi(\frac{-(D-6)(D-2)}{D-4}+24 \xi+48(D-4)\xi^2)} \\
\phi_1&=&-\frac{2(r_h b_1-1)}{r_h^2b_1(D-4)(1+4\xi)}
\end{eqnarray*}
where $\xi=\alpha/r_h^2$.  For all $D\ge5$ the discriminant is positive,
\begin{equation*}
(D-2)^2-48(D-2)^2\xi^2-1152(D-4)\xi^3 \geq 0,
\end{equation*}
when $\xi<\xi_c$, where $\xi_c$ is some (dimension dependent) positive critical value of $\xi$ and negative at
$\xi>\xi_c$ (above $\xi_c$ no real solutions exist). Then the lower bound of the radius of the horizon is at
$r_h=\sqrt{\alpha/\xi_c}$.  In particular, at $D=5$ $\xi_c=1/8$, in agreement with KT.

To obtain information about the mass gap we need to associate the radius of the horizon with the mass.
In the absence of exact solutions, this can only be done with an expansion that converges both at the horizon and at infinity: the small $\alpha$ expansion.  Since the maximal value
of $\alpha$, relevant for black brane solutions is obtained from $0.123<\xi_c<0.145$ for all $D>4$, it is likely that the expansion has favourable convergence properties.

Combining this result with the series expansion of the next subsection we obtain a result for the minimal mass
of black branes at $D\ge 5$. This will allow us to investigate the thermodynamics and phases of these black objects.

\subsection{Small $\alpha$ expansion}

In the previous section we obtained an expression for the minimal radius of horizon of black branes.
As shown by KT, at $D=5$, such a minimum  leads to a maximal value of the of the parameter $\sqrt{\alpha}/G_4M$
for black strings, where $M$ is a mass parameter. We restrict ourselves to black branes with horizons,
therefore it is natural to expect that black branes also satisfy similar bounds.  Furthermore, as we do
not employ numerical techniques, we restrict ourselves to an expansion of the metric components into
a power series of the dimensionless parameter $\beta=\alpha L^{2(D-4)}/(G_D M)^2$ at fixed variable
$\rho=rL^{D-4}/(G_D M)$. As the range of $\beta$ is bounded from above, we have a chance of obtaining
useful information from such an expansion.

We will use the ansatz
\begin{eqnarray}
g_{tt}=F_0(\rho)+\beta F_1(\rho)+\beta^2F_2(\rho)+...~,\nonumber\\
g_{rr}=G_0(\rho)+\beta G_1(\rho)+\beta^2G_2(\rho)+...~,\nonumber\\
g_{ww}=1+\beta H_1(\rho)+\beta^2H_2(\rho)+...~.
\label{beta-expansion}
\end{eqnarray}
Comparing with (\ref{metric1}) provides expressions for the expansion of functions $A(r),~B(r),$ and
$\phi(r)$, in terms of $F_i(\rho)$, $G_i(\rho)$ and $H_i(\rho)$, respectively, which can than be
substituted into (\ref{einstein2}) and (\ref{lanczos2}).  Then (\ref{einstein}) gives linear
differential equations for the expansion coefficients that can be solved order by order.  The solutions
of these equations for the first few expansion coefficients are listed in Appendix B.

Then (\ref{beta-expansion}) can be used to determine the radius of the event horizon in $\alpha$-expansion.
The horizon radius can be expressed as
\begin{equation}
r_h=\frac{2G_DM}{L^{D-4}}f(\beta)
\end{equation}
where $f(\beta)$ can be written as a power series in $\beta$ and $f(0)=1$. The horizon radius to 5th
order in $\beta=\alpha L^{2(D-4)}/(G_DM)^2$ for $D=6$
\begin{equation}
r_h=\frac{2G_6 M}{L^2}\left(1-\frac{23}{24}\beta-\frac{9479}{7200}\beta^2-\frac{5127979}{1505280}\beta^3-\frac{32357632419157}
{2950649856000}\beta^4-\frac{84365429160006541 }{2137113538560000}\beta^5+...\right).
\end{equation}
Using the lower bound obtained for the radius of the horizon we can derive lower bounds for the mass parameter,
$M$. In particular, we obtain the fifth order result for $D=6$ $G_6M_0/L^2=2.15~\alpha^{1/2}$.  The third order result
for $D=5$  agrees to within 4.4\% with the numerical value $G_6M_0/L^2=1.9843\alpha^{1/2}$ obtained in KT.

As we mentioned earlier, the asymptotic charge $Q$ has a unique relationship with $M$.
The non-vanishing value of $Q$ is related to the relative tension introduced in the study of black holes in
compactified Einstein gravity~\cite{harmark,kol}. $Q$ can also be obtained in the $\alpha$ expansion using
equation (\ref{beta-expansion}) and the expression for $H_i(\rho)$ given in Appendix B.

\subsection{Thermodynamics of Membranes}

Following Myers and Simon~\cite{myers} we assume that the relationship between the temperature and the surface gravity is the same as in Einstein gravity. Therefore we have
\begin{equation}
T=\frac{\kappa}{2\pi},
\label{temperature2}
\end{equation}
where
\begin{equation}
\kappa^2= -\frac{1}{2} (\nabla_a\xi_b)^2\vert_{\mathcal H},
\label{kappa}
\end{equation}
where $\xi_a$ is a normalized time-like Killing vector field and the right hand side is evaluated on
the Killing horizon, $\mathcal H$. Using metric (\ref{metric1}) the surface gravity is
\begin{equation}
\kappa=A'e^{A-B}.
\label{surface}
\end{equation}
The surface gravity can easily be calculated using our $\alpha$-expansion.  For general $D$ we obtain
in fourth order the expression,
\begin{eqnarray}
\kappa&=&\frac{L^{D-4}}{4G_DM}\left(1+\frac{D-4}{D-2}\beta+\frac{(D-4)(459 D-1690)}{120(D-2)^2}\beta^2+
\frac{13(23363D-81112)(D-4)^2}{15120(D-2)^3}\beta^3\right.\nonumber\\
&+&\left. \frac{(D-4)(-428992117312 + 344335921848 D - 91865320338D^2 + 8150296327D^3)}
{66528000(D-2)^4}\beta^4\right)
\end{eqnarray}
while for $D=6$ in fifth order
\begin{equation}
\kappa=\frac{L^2}{4G_6M}\sum_{k=0}^\infty c_k \beta^k=\frac{L^2}{4G_6M}\left(1+\frac{1}{2}\beta+
\frac{133}{120}\beta^2+\frac{54847}{17280}\beta^3+\frac{1129198603}{106444800}\beta^4+
\frac{30955288943}{3193344000}\beta^5+...\right)
\label{kappa6}
\end{equation}
where $\beta=\alpha L^4/(G_6M)^2$.

It is instructive to investigate the convergence properties of (\ref{kappa6}). Assuming a finite
radius of convergence we analyze the coefficients to find the radius of convergence and critical
behavior.  We choose the following simple ansatz for the coefficients
\begin{equation}
c_{k}/c_{k-1}=c(1+\frac{\delta}{k}).
\end{equation}
Using linear regression we  find the following values
\begin{equation}
c=4.27\pm0.12,~~~~~~\delta=-3.8\pm0.2
\end{equation}
The value of the critical exponent $\delta<-1$ seems to indicate that the series converges at the
critical point. The critical value of $M$ is given by $G_6M_0/L^2=(2.07\pm 0.03)\sqrt{\alpha}.$ This is in
a excellent agreement with the critical value obtained from the horizon expansion, $G_6M_0/L^2=2.15
\sqrt{\alpha}$.  The temperature is singular at the minimal value of mass, but it seems to stay
finite at this point.  It behaves like
\begin{equation}
T-T_c\sim \left(\frac{G_6M}{L^2}-\sqrt{c\alpha}\right)^{\delta-1}
\label{critical}
\end{equation}
Note that for integer $\delta$ an extra multiplier log$(G_6M/L^2-\sqrt{c\alpha})$ appears in (\ref{critical}).
Though $T$ is finite at the critical mass, it becomes complex below this value, implying that the
radius of the horizon is complex. Having no real solution for the radius of horizon is equivalent
to not having an event horizon: a naked singularity. Naked singularities are expected to be unstable~ \cite{vaz}.

The first law of thermodynamics for D-dimensional black membranes with D-4 compactiifed dimensions reads \cite{traschen,harmark,kol,harmark2}
\begin{equation}
dS= \frac{dM_\text{ADM}}{T} - \sum_{i=1}^{D-4}\frac{\tau_i}{T} dL_i
\end{equation}
where $M_\text{ADM}$ is the ADM mass. For the case under consideration
\begin{equation}\label{definemadm}
M_\text{ADM}=\frac{V_{D-4}\Omega_2}{16\pi G_D}\left[2c_t-\sum_{i=1}^{D-4} c_i\right]=
\frac{L^{D-4}}{4G_D}\left[2c_t-(D-4) c_w\right],
\end{equation}
and $\tau_i$ are tensions along the toroidal extra dimensions and are given by,
\begin{equation}\label{definetau}
\tau_i = \frac{V_{D-4} \Omega_2}{16\pi G_D L_i}\left[c_t - 2c_i -\sum_{j\neq i} c_i\right]
= \frac{L^{D-5}}{4G_D}\left[c_t-(D-3)c_w\right]
\end{equation}
where $c_i=c_w$ for all $i$ and $c_t$ and $c_w$ are leading order corrections to the metric
\begin{eqnarray}
g_{tt}&=&\eta_{tt}+h_{tt}=-1+\frac{c_t}{r},\cr\cr
g_{ww}&=&\eta_{ww}+h_{ww}=1+\frac{c_w}{r},
\end{eqnarray}
and $V_{D-4}$ is the volume of the compact dimensions.
Using the formul\ae\ given in Appendix B, we find, up to second order in $\beta$
\begin{eqnarray}\label{ctcw}
c_t&=&\frac{2 G_D M}{L^{D-4}},\cr
c_w&=&\frac{2 G_D M}{L^{D-4}}\left(\frac{2}{D-2}\beta+\frac{133(D-4)}{30(D-2)^2}\beta^2+\cdots\right).
\end{eqnarray}
We use the above expresions in (\ref{definemadm}) and (\ref{definetau}) to calculate
\begin{eqnarray}\label{madm}
M_\text{ADM} = M\left[1- (D-4)\left(\frac{\beta}{D-2} + \frac{(1995D-7980)}{900(D-2)^2}\beta^2+...\right)\right]
\end{eqnarray}
and
\begin{equation}\label{tau}
\tau = \frac ML \left[\frac 12 - (D-3)\left(\frac{\beta}{D-2} + \frac{(1995D-7980)}{900(D-2)^2}\beta^2+...\right)\right].
\end{equation}
From here we calculate
\begin{eqnarray}
\frac{1}{T}&=&\frac{8\pi G_D M_\text{ADM}}{L^{D-4}}\left[1-\frac{73}{120}\left(\frac{D-4}{D-2}\right)\beta'^2+\cdots\right]\cr\cr
\frac{\tau}{T}&=&\frac{4\pi G_D M_\text{ADM}^2}{L^{D-3}}\left[1-\beta'-\frac{73}{40}\left(\frac{D-4}{D-2}\right)\beta'^2
+\cdots\right].
\end{eqnarray}
where $\beta'=\alpha L^{2(D-4)}/(G_D M_\text{ADM})^2$. With these
equations we verify that the integrability conditions
\begin{equation}
\frac{\partial}{\partial L}\left(\frac 1T \right)_{M_\text{ADM}} = -\frac{\partial}{\partial M_\text{ADM}}
\left(\frac{\tau}{T}\right)_L
\label{integrability}
\end{equation}
are satisfied.

Having an analytic expression for the temperature as a function of the ADM mass allows us to calculate the
entropy, up to an undetermined constant $S_0$
\begin{equation}
S(M_\text{ADM},L)=\int \frac{\partial M_\text{ADM}}{T} - \int \left[ \sum_{i=1}^{D-4}\frac{\tau_i}{T} +\frac
{\partial}{\partial L} \int \frac {\partial M_\text{ADM}}{T}\right] dL + S_0
\end{equation}
For general $D$ up to second order in $\beta'$, we obtain
\begin{equation}
S=\frac{4\pi G_DM_{\text{ADM}}^2}{L^{D-4}}\left(1+\beta' + \frac{73(D-4)}{120(D-2)}\beta'^2+...
\right) + S_0,
\label{entstr}
\end{equation}
which reduces in $D=6$ to the following expression, up to a constant term independent of $M_\text{ADM}$
and $L$,
\begin{equation}
S_6=\frac{4\pi G_DM_{\text{ADM}}^2}{L^2}\left(1+\beta' + \frac{73}{240}\beta'^2+...\right)
\label{entstr2}
\end{equation}
Fursaev and Solodukhin introduced a geometric approach entropy in \cite{fursaev}. We quote their result below,
\begin{equation}
S=\frac{A_\Sigma}{4G_D}-\int_\Sigma \left(8\pi c_1 R + 4\pi c_2 R_{\mu\nu} n^\mu_i n^\nu_i
+8\pi c_3 R_{\mu\nu\lambda \rho} n^\mu_i n^\lambda_i n^\nu_j n^\rho_j\right)
\label{fursaev}
\end{equation}
where $c_1 = \alpha/16\pi G_D$, $c_2 = -4\alpha/16\pi G_D$ and $c_3 = \alpha/16\pi G_D$ for EGB gravity,
and $n_i^\mu$ are the two orthonormal vectors orthogonal to $\Sigma$, where $\Sigma$ is
the horizon surface. The expression for entropy (\ref{entstr2}) agrees with that calculated from
(\ref{fursaev}).

\section{Comparison of entropies of black holes and black branes}
The horizon radius $r_h$ of the spherical black hole in $D$ dimensions is obtained as the real, positive root
of the equation $f(r_h)=0$, or
\begin{equation}
r_h^{D-3} + \tal r_h^{D-5} - \tM =0,
\end{equation}
where $\tal$ and $\tM$ are defined in (\ref{talpha}) and (\ref{tM}).
This equation cannot be solved exactly for general $D$, but for $D=6$ one finds
\begin{equation}
r_h=\frac{2^{4/3}}{(\tM+\sqrt{\tM^2+32\tal^3})^{1/3}}+\frac{(\tM+\sqrt{\tM^2+32\tal^3})^{1/3}}{2^{1/3}}.
\end{equation}
These formul\ae~ will no longer hold when we compactify $D-4$ dimensions. However when the horizon radius of the black hole is smaller than the compactification radius we expect the corrections to be relatively small. In a previous study of black holes in Einstein gravity compactified on $S^1$ we observed that the corrections are of the order of 10\% \cite{karasik}.

In 6 dimensions
\begin{equation}
\tM = \frac{3G_6M}{2\pi}.
\end{equation}
The exact expression for the entropy is not illuminating, but it is worth noting that its leading
behavior is as $M_\text{adm}^{4/3}$ (for large $M_\text{adm}$) so that it grows much slower than $S_6$ for the black
membrane in (\ref{entstr2}), whose leading behavior is $M_\text{adm}^2$. In FIG 1, we plot the black hole
entropy (blue, solid) and the string entropy (red, dashed), taking $G_6/L^2=1$, $L=3$ and $\alpha = 0.1 $ in the
figure on the left, and $G_6/L^2=1$, $L=10$ and $\alpha = 0.1$ in the figure on the right. The expansion
used to obtain the membrane entropy is valid only when $\beta' \ll 1$. Therefore, if we let
$\alpha=a G_6/L^2$ ($a$ is dimensionless), it is valid only for masses $M_\text{adm} \gg \sqrt{\alpha}
\sim 0.316 L/\sqrt{G_6}$ for our choice of $\alpha$. From the graphs, the entropy of the black membrane is seen
to fall below the entropy of the 6 dimensional black hole within the range of validity of the $\beta$
expansion, if $L$ is chosen to be sufficiently large (in the particular case of $\alpha = 0.1$,
$L> 2.376$ Planck lengths). The mass at crossover grows as $0.133 L^3/G_6$ to leading order.

\begin{figure}[t]
\setlength{\unitlength}{1cm}
\begin{minipage}[t]{7.5cm}
\begin{center}
\begin{picture}(7.5,5.0)
\epsfig{file=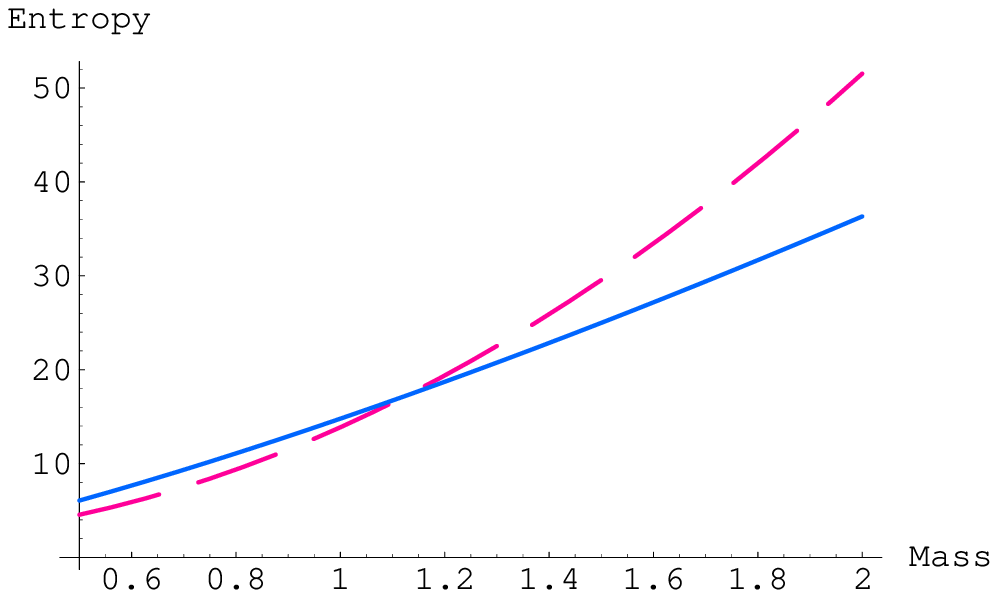,width=\hsize}
\end{picture}\par
\end{center}
\end{minipage}
\hfill
\begin{minipage}[t]{7.5cm}
\begin{center}
\begin{picture}(7.5,5.0)
\epsfig{file=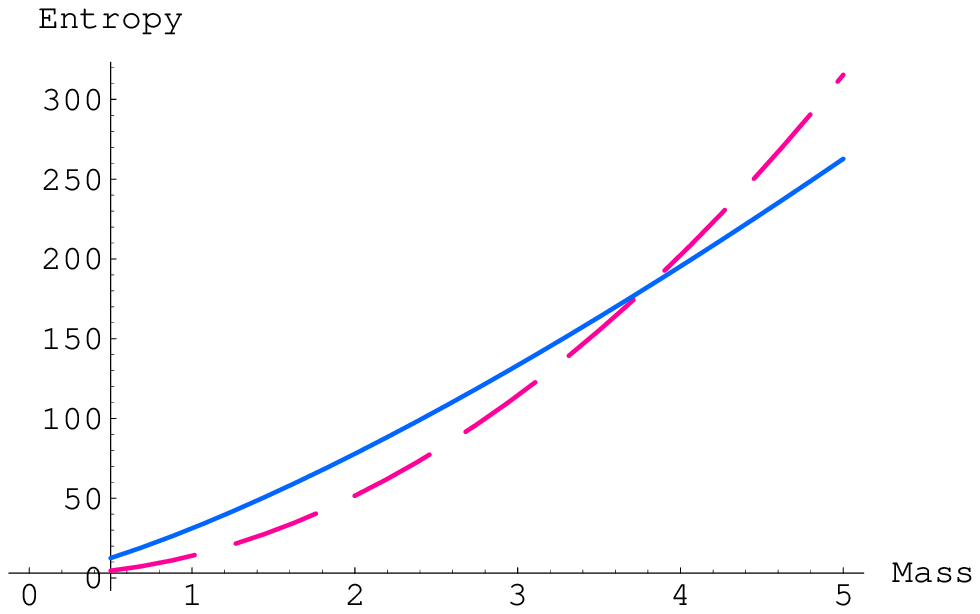,width=\hsize}
\end{picture}\par
\end{center}
\end{minipage}
\caption{\small Plots of the entropy of the 6 dimensional black hole (blue, solid) and the black membrane (red,
dashed) with compactification on a two torus. In the first figure, we have taken $\alpha=0.1$,
$G_6/L^2=1$ and $L=3$ in Planck units. In the second figure, $L=10$. The entropy of the black membrane falls
below the entropy of the black hole at some critical mass which depends on the length scale of the
extra dimensions.}
\label{fig:Vppn}
\end{figure}

\section{summary}
In this paper we have analyzed black brane solutions in D-dimensional Einstein Gauss-Bonnet gravity
compactified on a D-4 torus. The metric cofficients are given as series expansions in terms of the
Gauss-Bonnet coupling parameter. A separate expansion about the horizon yields lower bound
of the horizon radius indicating that there is a mass gap for Gauss-Bonnet black branes. We have calculated
the entropies of such objects and compared them with that of D dimensional black-holes in two
independent ways . A comparison of the entropy of the black membrane and the entropy of the black hole
indicates that there is a critical mass at which the entropy of the black membrane is smaller than
that of the black hole. The critical mass depends on the radius of compactification.

\begin{acknowledgments}
This work was supported in part by the U.S. Department of Energy Grant No. DE-FG02-84ER40153. We thank Philip Argyres
for useful discussions. L.C.R.W wishes
to thank the Aspen Centre for Physics for hospitality during the summer of 2005, and the
participants of the workshop on the physics of black holes for valuable discussions.
\end{acknowledgments}

\appendix
\section{Einstein and Lanczos tensors in $D$ dimensions}

The generalization of the expressions for the Einstein and Lanczos tensors of Kobayashi and Tanaka are given by
\begin{eqnarray}
G_t^t&=&\frac{e^{-2B}}{r^2}\{1-e^{2B}-2rB'+(D-4)[2r\phi'+r^2(\phi')^2\frac{D-3}{2}- r^2B'\phi'+r^2\phi'' ]\},\nonumber\\
G_r^r&=&\frac{e^{-2B}}{r^2}\{1-e^{2B}+2rA'+(D-4)[2r\phi'+r^2(\phi')^2\frac{D-5}{2}+ r^2A'\phi']\}, \nonumber\\
G_\theta^\theta&=&\frac{e^{-2B}}{r}\{(1+rA')(A'-B')+(D-4)[\phi'+r(\phi')^2\frac{D-3}{2}+r\phi'(A'-B')+r\phi'' ]\},
\nonumber\\
G_w^w&=&\frac{e^{-2B}}{r^2}\{1-e^{2B}+r(A'-B')(2+rA')+r^2A''+(D-5)r[2\phi'+r\phi'(\phi'\frac{D-4}{2}+A'-B')+r\phi'' ]\},
\nonumber\\
\label{einstein2}
\end{eqnarray}
\begin{eqnarray}
H_t^t&=&\frac{8(D-4)e^{-2B}}{r^2}\left\{(e^{2B}-3)B'\phi'-(e^{2B}-1)\phi''-\frac{\phi'^2}{2}[(D-3)e^{2B}-3D+13]+(D-5)\right.
\nonumber\\
&\times&\left.\left[\frac{1}{8}(D-3)(D-6) r^2\phi'^4-\frac{D-6}{2}r^2(B'\phi'^3-\phi'^2\phi'')-3rB'\phi'^2+(D-4)r\phi'^3+2r
\phi'\phi''\right]\right\},\nonumber\\
H_r^r&=&\frac{8(D-4)e^{-2B}}{r^2}\left\{(3-e^{2B})A'\phi'+\frac{D-5}{2}\phi'^2\right.\nonumber\\
&\times&\left.\left[6rA'+3-e^{2B}+(D-6)r^2\phi'A'+2(D-6)r\phi'+\frac{(D-6)(D-7)}{4}r^2\phi'^2
\right]\right\},\nonumber\\
H_\theta^\theta&=&\frac{8(D-4)e^{-2B}}{r^2}\left\{A'\phi''+\phi'(A'^2-3A'B'+A'')+\frac{3D-13}{2}a'\phi'^2+(D-5)\right.
\nonumber\\
&\times&\left.\left[rA'\phi\phi''+\frac{D-6}{2}r\phi''\phi'^2-\frac{3}{2}rA'B'\phi'^2+\frac{1}{2}(rA'^2+rA''-3B')\phi'^2
\right.\right.\nonumber\\&+&\left. \left. \frac{D-4}{2}(1+rA')\phi'^3-\frac{D-6}{2}rB'\phi'^3+\frac{(D-3)(D-6)}{8}r\phi'^4
\right]\right\},\nonumber\\
H_w^w&=&\frac{8e^{-4B}}{r^{2}}\left\{\left(1-e^{2B}\right)A'^2\left(-3+e^{2B}\right)A'B'+\left(1-e^{2B}\right)A''+(D-5)\right.
\nonumber\\&\times&\left. \left[1-e^{2B}+2rA'+(D-6)(2r+r^2A')\phi'+\frac{1}{2}(D-6)(D-7)r^2\phi'^2\right]\phi''+(D-5)\phi'
\right.\nonumber\\&\times&\left.\left[2rA'^2+(-3+e^{2B})B'+A'(3-e^{2B}-6rB')+2rA''\right]+\phi'^2
\right.\nonumber\\&\times&\left. \left[-\frac{1}{2}(D-4)(D-5)e^{2B}+\frac{1}{2}(3D-16)(D-5)(1+2rA')
\right.\right.\nonumber\\&+&\left. \left.
\frac{1}{2}(D-6)(D-5)(r^2A'^2+r^2A''-6rB'-3r^2A'B')\right]+\phi'^3\right. \nonumber\\&\times&\left.\left[\frac{1}{2}
(D-5)^2(D-6)(2r+r^2A')-\frac{1}{2}(D-7)(D-5)(D-6)r^2B'\right]+\frac{1}{8}(D-7)(D-6)(D-5)(D-4)r^2\phi'^4    \right\}
\nonumber \\
\label{lanczos2}
\end{eqnarray}

\section{Coefficients of the $\alpha$-expansion of the components of the metric tensor}
The coefficients of (\ref{beta-expansion}) up to fourth order in $\beta=\alpha L^{2(D-4)}/(G_DM)^{2}$ are
\begin{eqnarray}
F_0(\rho)&=&1-\frac{2}{\rho},\nonumber\\
F_1(\rho)&=&\frac{2(D-4)(6\rho^2+5\rho+12)}{3(D-2)\rho^4},\nonumber\\
F_2(\rho)&=&-\frac{2400\left(13D^{2}-94D+168\right)+32\left(8864+-4252D+509D^{2}\right)\rho+36\left(8904-4342D+529D^{2}\right)
\rho^{2}}{450\left(D-2\right)^{2}\rho^{7}}\nonumber \\
&& +\frac{390\left(19D^{2}-172D+384\right)\rho^{3}+25\left(71D^{2}-580D+1184\right)\rho^{4}}{450\left(D-2\right)^{2}
\rho^{7}}\nonumber\\
F_3(\rho)&=&\frac{125400\left(213D^{3}-2294D^{2}+8020D-9008\right)+2352\left(60189D^{3}-692228D^{2}+2646352D-3361856\right)
\rho}{264600\left(D-2\right)^{3}\rho^{10}}\nonumber\\&&+\frac{720\left(211835D^{3}-2481484D^{2}+9687152D-12602304\right)
\rho^{2}}{264600\left(D-2\right)^{3}\rho^{10}}\nonumber\\&&+\frac{48\left(1425559D^{3}-18065358D^{2}+77255232D-111210976
\right)\rho^{3}}{264600\left(D-2\right)^{3}\rho^{10}}\nonumber\\&&+\frac{224\left(132535D^{3}-1747248D^{2}+7654104D-11142688
\right)\rho^{4}}{264600\left(D-2\right)^{3}\rho^{10}}\nonumber\\&&-\frac{2646\left(4473D^{3}-47280D^{2}+161936D-177536\right)
\rho^{5}}{264600\left(D-2\right)^{3}\rho^{10}}\nonumber\\&&-\frac{210\left(44287D^{3}-562832D^{2}+2397040D-3417216\right)
\rho^{6}}{264600\left(D-2\right)^{3}\rho^{10}}\nonumber\\&&-\frac{35\left(D-4\right)^{2}\left(128923D-568696\right)\rho^{7}
+210\left(D-4\right)^{2}\left(50185D-194608\right)\rho^{8}}{264600\left(D-2\right)^{3}\rho^{10}}\nonumber\\
G_0(\rho)&=&\frac{1}{1-\frac{2}{\rho}},\nonumber\\
G_1(\rho)&=&-\frac{2(D-4)(6\rho^3+3\rho^2+3\rho-20)}{3(D-2)\rho^2(\rho-2)^2},\nonumber\\
G_2(\rho)&=&-\frac{1600\left(7D^{2}+82D-440\right)-96\left(403D^{2}-2014D+1608\right)\rho+24\left(511D^{2}-4118D+8296\right)
\rho^2}{450\left(D-2\right)^{2}\left(\rho-2\right)^{3}\rho^{5}}\nonumber\\&&+\frac{40\left(428D^{2}-4159D+9788\right)
\rho^{3}+10\left(1067D^{2}-7156D+11552\right)\rho^{4}-15\left(673D^{2}-5324D+10528\right)\rho^{5}}{450\left(D-2\right)^{2}
\left(\rho-2\right)^{3}\rho^{5}}\nonumber \\&&-\frac{45\left(313D^{2}-2524D+5088\right)\rho^{6}+3990\left(D-4\right)^{2}
\rho^{7}}{450\left(D-2\right)^{2}\left(\rho-2\right)^{3}\rho^{5}}\nonumber
\label{coefficients}
\end{eqnarray}
\pagebreak
\begin{eqnarray}
G_3(\rho)&=&\frac{10035200\left(264D^{3}-2885D^{2}+10318D-12008\right)}{264600\left(D-3\right)^{3}\left(\rho-2\right)^{4}
\rho^{8}}\nonumber\\&&-\frac{15680\left(193721D^{3}-2015348D^{2}+6727696D-7063360\right)\rho}{264600\left(D-3\right)^{3}
\left(\rho-2\right)^{4}\rho^{8}}\nonumber\\&&+\frac{64\left(15179233D^{3}-152626756D^{2}+483893264D-465015872\right)
\rho^{2}}{264600\left(D-3\right)^{3}\left(\rho-2\right)^{4}\rho^{8}}\nonumber\\
&&-\frac{16\left(25877281D^{3}-313845372D^{2}+1256103888D-1659035584\right)\rho^{3}}{264600\left(D-3\right)^{3}
\left(\rho-2\right)^{4}\rho^{8}}\nonumber\\
&&+\frac{512\left(744343D^{3}-7380048D^{2}+22593720D-19932064\right)\rho^{4}}{264600\left(D-3\right)^{3}\left(\rho-
2\right)^{4}\rho^{8}}\nonumber\\
&&-\frac{168\left(936799D^{3}-8324156D^{2}+20820496D-10050624\right)\rho^{5}}{264600\left(D-3\right)^{3}\left(\rho-
2\right)^{4}\rho^{8}}\nonumber\\
&&+\frac{56\left(2020247D^{3}-27338280D^{2}+121583184D-178216064\right)\rho^{6}}{264600\left(D-3\right)^{3}\left(\rho-
2\right)^{4}\rho^{8}}\nonumber\\
&&+\frac{70\left(383801D^{3}-3246576D^{2}+7852560D-4028288\right)\rho^{7}}{264600\left(D-3\right)^{3}\left(\rho-2
\right)^{4}\rho^{8}}\nonumber\\
&&-\frac{140\left(520333D^{3}-6146976D^{2}+24260304D-31990912\right)\rho^{8}}{264600\left(D-3\right)^{3}\left(\rho-
2\right)^{4}\rho^{8}}\nonumber\\
&&-\frac{105\left(D-4\right)^{2}\left(663701D-2741576\right)\rho^{9}+735\left(D-4\right)^{2}\left(78913D-312712
\right)\rho^{10}}{264600\left(D-3\right)^{3}\left(\rho-2\right)^{4}\rho^{8}}\nonumber\\
&&-\frac{210\left(D-4\right)^{2}\left(50185D-194608\right)\rho^{11}}{264600\left(D-3\right)^{3}\left(\rho-2
\right)^{4}\rho^{8}}\nonumber\\
H_1(\rho)&=&\frac{4(3\rho^2+3\rho+4)}{3(D-2)\rho^3},\nonumber\\
H_2(\rho)&=&\frac{4\left(-1995\rho^{5}+255\rho^{4}+3340\rho^{3}+6060\rho^{2}+5856\rho+4800\right)}{225\left(D-2
\right)^{2}\rho^{6}}\nonumber\\&&+\frac{D\left(1995\rho^{5}+195\rho^{4}-2440\rho^{3}-4410\rho^{2}-4656\rho-4000
\right)}{225\left(D-2\right)^{2}\rho^{6}}\nonumber \\
H_3(\rho)&=&\frac{627200\left(269D^{2}-1894D+3072\right)+7056\left(15087D^{2}-120464D+236464\right)\rho}{396900\left
(D-2\right)^{2}\rho^{9}}\nonumber \\
&&+\frac{576\left(137595D^{2}-1242076D+2810184\right)+336\left(75071D^{2}-928154D+2776080\right)\rho^{3}}{396900
\left(D-2\right)^{2}\rho^{9}}\nonumber\\
&&+\frac{2016\left(4243D^{2}-59508D+197444\right)\rho^{4}-630\left(15951D^{2}-135028D+244576\right)\rho^{5}}
{396900\left(D-2\right)^{2}\rho^{9}}\nonumber\\
&&-\frac{420\left(23357D^{2}-228520D+530288\right)\rho^{6}+315\left(5497D^{2}+6844D-115328\right)\rho^{7}}
{396900\left(D-2\right)^{2}\rho^{9}}\nonumber\\
&&+\frac{315\left(50185D^{2}-395348D+778432\right)\rho^{8}}{396900\left(D-2\right)^{2}\rho^{9}}\nonumber\\
\end{eqnarray}
where we have used the notation $\rho=r L^{D-4}/(G_D M)$.

\end{document}